\documentclass[a4paper,PRL,preprint,showpacs, showkeys]{revtex4}
\usepackage{graphicx}
\usepackage{flafter}
\usepackage{epsfig}
\usepackage{color}
\usepackage{amssymb}
\usepackage{gensymb}
\usepackage[top=2.5cm, bottom=2.5cm, left=1.8cm, right=1.8cm]{geometry}
\usepackage{color,soul}
\begin{document}

\newcommand\igorfig[1]{\includegraphics[scale=0.8]{#1}}
\renewcommand{\topfraction}{.9}
\renewcommand{\bottomfraction}{.3}
\renewcommand{\textfraction}{.1}
\renewcommand{\floatpagefraction}{.7}
\renewcommand{\dbltopfraction}{.75}
\renewcommand{\dblfloatpagefraction}{.85}
\setcounter{topnumber}{3}
\setcounter{bottomnumber}{2}
\setcounter{totalnumber}{20}
\setcounter{dbltopnumber}{3}

\newcommand\mapfigure[3]{\begin{figure}
            \begin{center}
            \includegraphics[scale=0.8]{#1}
            \end{center}
            \caption{\label{#3}#2}
            \end{figure}
            }


\title{Origin of C$_{60}$ surface reconstruction resolved by atomic force microscopy }
\author{Leonardo Forcieri$^1$}
\author{Simon Taylor$^2$}
\author{Philip Moriarty$^2$}
\author{Samuel P Jarvis$^1$}

\affiliation{$^1$ Physics Department, Lancaster University, Lancaster, LA1 4YB, UK}
\affiliation{$^2$ The School of Physics and Astronomy, The University of Nottingham, Nottingham NG7 2RD, UK}

%


\begin{abstract}
Surface adsorption of C$_{60}$ affects its chemical and electronic properties.  Numerous studies have reported observation of bright and dark fullerenes on metal surfaces that suggest extensive surface reconstruction, however, the underpinning mechanism of the reconstruction remains under debate.  Here we report tip-functionalised non-contact atomic force microscope (ncAFM) measurements which unambiguously reveal that C$_{60}$ fullerenes adsorb with three well-defined adsorption heights on the Cu(111) surface, consistent with theoretical reports of top-layer hollow site, single-atom vacancies, and surface nanopits.  Using single molecule resolution $\Delta f(z)$ measurements we identify well defined adsorption heights specific to each site, confirming the presence of a complex vacancy model for C$_{60}$ monolayers on metal surfaces.

\end{abstract}

\maketitle



Adsorption of C$_{60}$ fullerenes on metal (111) surfaces is a surprisingly complex problem. The propensity for C$_{60}$ to adopt an optimum nearest neighbour distance results in a rich variety of surface interactions\cite{Gardener2009,Li2009,Pai2010,Moriarty2010}, in some cases resulting in extensive and long-range reconstructions of the C$_{60}$ layer\cite{Pai2004,Schull2007}. The origin of this apparent surface reordering, and the existence, or lack thereof, of a vacancy reconstruction has been intensely debated for decades.  Scanning tunnelling microscopy (STM) has provided valuable insight into this highly substrate dependent\cite{Ledieu2017} problem, identifying patterns of bright and dark molecules across many surfaces including Au(111)\cite{Gardener2009,Shin2014,Tang2011,Chandler2019}, Ag(111)\cite{Colton1993}, Cu(111)\cite{Pai2010,Pai2004} and Pt(111)\cite{Pinardi2014}, with a multitude of models put forward to explain such observations.  It therefore remains unclear to what extent C$_{60}$ fullerenes induce surface vacancies, which not only affects surface geometry but also substantially modifies the surface interaction and electronic properties of fullerene systems. 

Several proposals for fullerene adsorption have been put forward.  On Au(111) both vacancy reconstruction\cite{Gardener2009} and hcp hollow site\cite{Wang2004} adsorption have been proposed, where rotational effects are thought to play a major role\cite{Leaf2016}.  On Ag(111), however, a compelling LEED I(V) analysis strongly supports the vacancy structure, suggesting a single atom is ejected under the C$_{60}$ fullerene\cite{Li2009}, furthermore, STM\cite{pussi2012} and x-ray standing wave\cite{Jarvis2021} studies suggest the existence of a mixed layer of vacancy and surface adsorbed fullerene. On surfaces with smaller surface lattice constants such as Cu(111) and Pt(111), even more complicated models have been put forward involving various stages of vacancy formation on Pt(111)\cite{Pinardi2014}, or the 7-atom nanopit model for Cu(111)\cite{Pai2010,Xu2012}.  Despite this wealth of information on C$_{60}$-metal systems, what remains lacking is direct experimental measurement of molecular height in order to unambiguously determine the adsorption state.

 \begin{figure}[!htbp]
	\includegraphics[width=8.5cm]{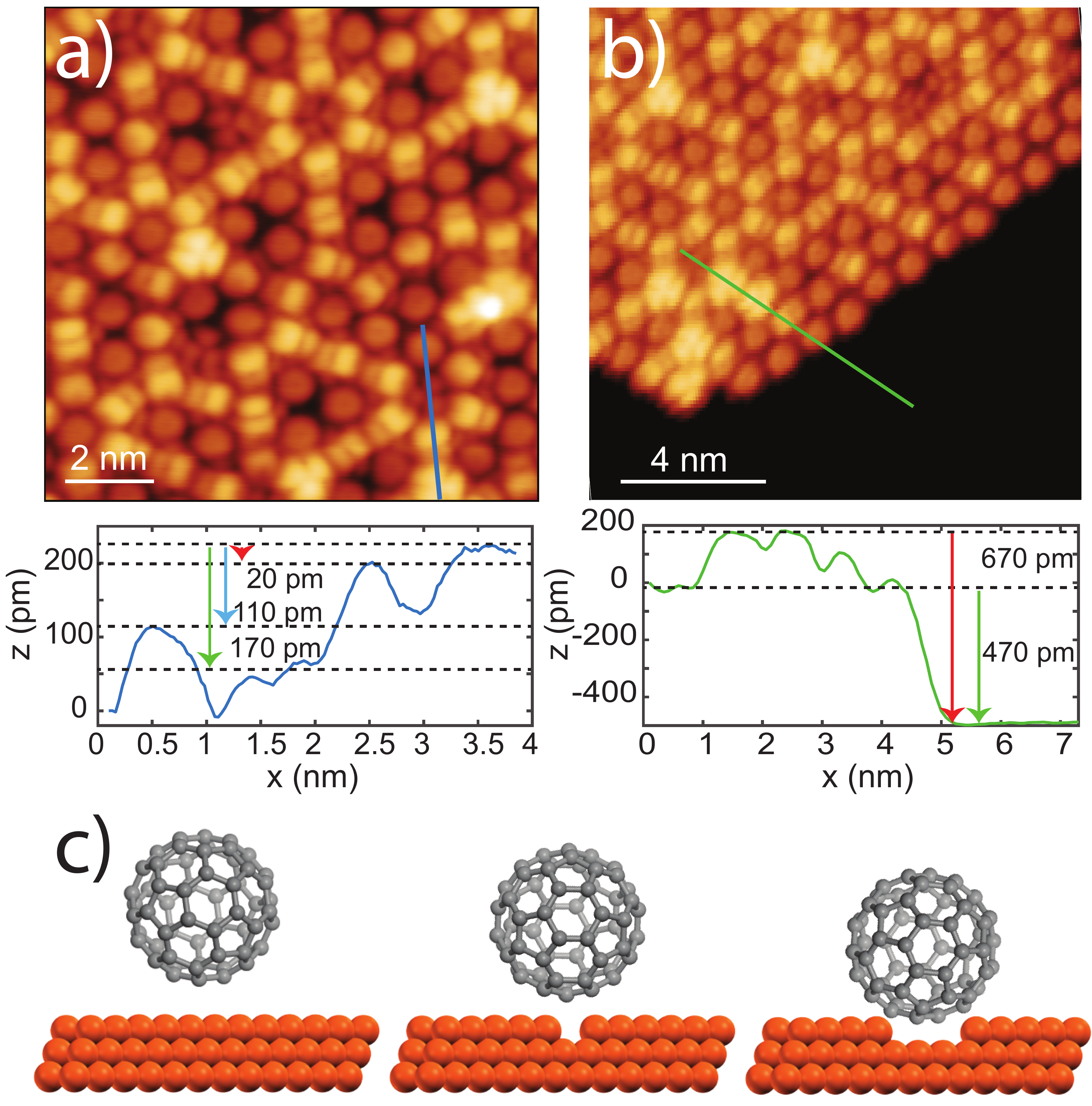}
	\caption{\label{fig:Fig1} \textbf{Room temperature reconstruction of Cu(111):C$_{60}$.} \textbf{(a)} Constant current submolecular resolution STM image showing the typical distribution of C$_{60}$ molecular brightness and orientation with respect to the Cu(111) surface.  \textbf{(b)} Constant current STM image at the edge of a molecular island.  Line profile analysis reveals four distinct heights within the C$_{60}$ island with respect to the Cu(111) surface. \textbf{(c)} Ball-and-stick cartoon depicting three possible adsorption configurations for the molecular heights observed.   Parameters: (a,b)+1.5 V/ 200 pA.  
	}
\end{figure}

Here we report measurements of the adsorption height of reconstructed C$_{60}$ fullerenes on a Cu(111) surface with submolecular resolution tip-functionalised STM and non-contact atomic force microscopy (ncAFM).  Short-range frequency shift, $\delta f(z)$, measurements with ncAFM reveal three distinct adsorption heights for the C$_{60}$ fullerenes consistent with theoretical studies reporting unreconstructed, vacancy site, and nanopit adsorption.  Furthermore, we find that the measured height variation is broadly consistent with STM height profiles, suggesting that STM contrast is almost entirely topographic in origin.  Finally, we show that the surface reconstruction is suppressed when samples are prepared at low temperatures such that no height variation is observed, confirming the reconstruction model, and ruling out significant contribution from molecular rotation.


Measurements were made on a Createc GmbH LT STM-AFM system controlled by Nanonis electronics.  C$_{60}$ fullerenes were deposited via standard thermal sublimation under ultrahigh vacuum conditions (better than 1 $\times 10^{-10}$ mbar) onto a clean sputter-annealed Cu(111) surface.  A commercial qPlus sensor (Createc GmbH) with a separate tunnel current wire was used for both the STM and ncAFM experiments (f$_0$ $\sim$ 20 kHz; Q$\sim$30,000 at 5 K; nominal spring constant 1,800 Nm$^{-1}$).




\textbf{Temperature dependence of surface reconstruction.} Figure \ref{fig:Fig1} shows STM images collected at 77 K for a sample prepared by depositing C$_{60}$ molecules onto a room temperature Cu(111) surface.  Similar to previous reports\cite{Pai2010,Pai2004}, clear variation in molecular appearance can be observed with characteristic dark and bright molecules visible within the molecular islands. 

Line profile analysis reveals four distinct molecular heights relative to the Cu(111) surface.  The measured heights are 670-700 pm for the two brightest observed molecules (two lines separated by the red arrow in Figure \ref{fig:Fig1}(a)), 470-500 pm for the darkest molecules (green arrow), and 570-600 pm for a configuration located in between (blue arrow).  These height variations are significantly beyond what can be attributed to molecular rotation, and appear more in-line with reports of vacancy and nanopit reconstruction models, which we schematically show in Figure \ref{fig:Fig1}(c).  Moreover, submolecular detail within the images reveal that certain adsorption heights have a preference for specific orientations, appearing as hexagon-up or C-C up orientations (bright - red), pentagon-up (middle - blue), or hexagon-up (dark - green).  Despite this apparent orientation preference, however, the observation that hexagon-up adopts two distinct heights suggests that rotation alone cannot fully explain the observed height variation\cite{Tang2011,Shin2014,Gardener2009,Pabens2015,Colton1993}.


\begin{figure}
	\includegraphics[width=6cm]{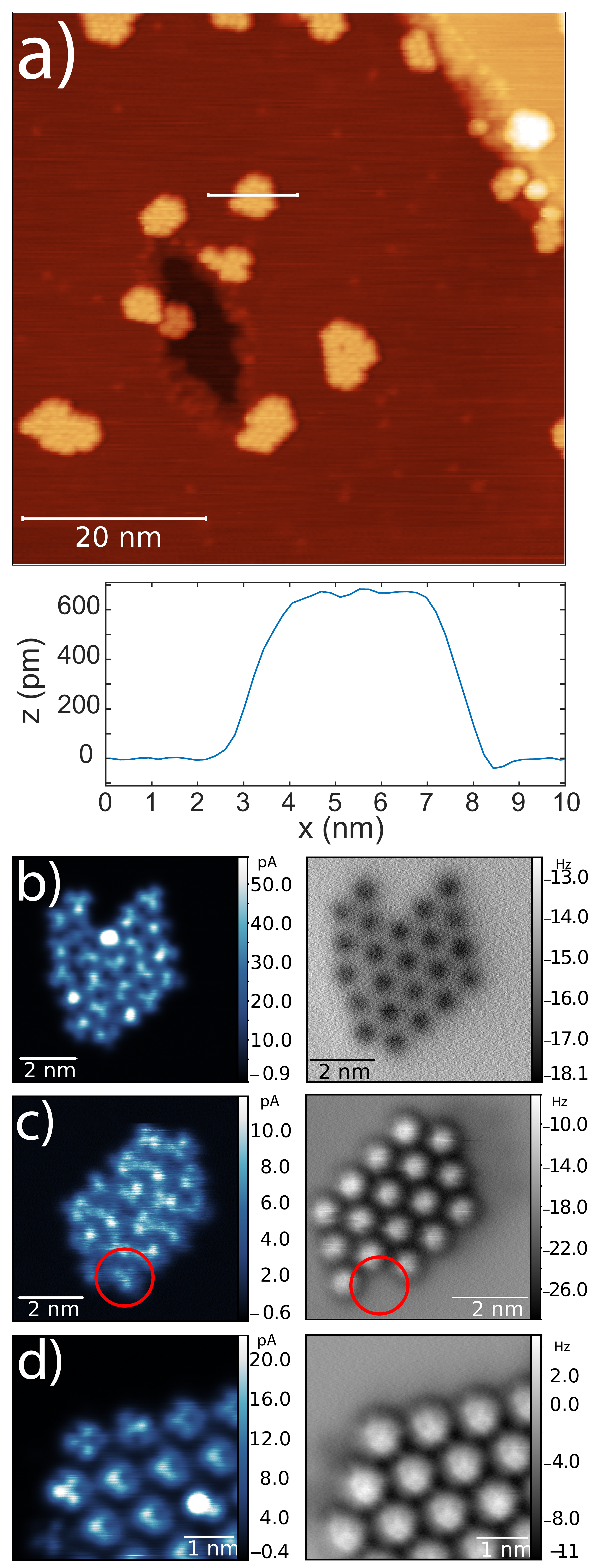}
	\caption{\label{fig:Fig2} \textbf{Suppressing surface reconstruction at low temperature.} \textbf{(a)} Constant current STM images of Cu(111):C$_{60}$ prepared at low temperatures shown with line profile measurements.  A measured island height of 670 $\pm$ 50 pm suggests no detectable surface reconstruction of the molecular island.    \textbf{(b-d)} Sequential constant height images of current (top) and frequency shift (bottom) for typical molecular islands.  Red circle in (c) shows an example removal of a C$_{60}$ used for tip-functionalisation.  Parameters: (a) Constant current images were collected in STM mode (i.e. no oscillation) with a sample bias of +1.5 V and set point current of 100 pA.  (b-d) For all constant height images an oscillation amplitude, a$_0$, of 300 pm was used.  Constant height current images used a sample bias of 100 mV.  For Constant height frequency shift images sample bias was varied until the tunnel current reached zero to eliminate the effect of cross-talk \mbox{\cite{Majzik2012}}; (b) V = -2.69 mV, (c) V = -1.9 mV, (d) V = -1.5 mV.   
		}
\end{figure}

In order to suppress the reconstruction of the underlying metal surface, a second set of samples was prepared by depositing C$_{60}$ fullerene onto the Cu(111) substrate held at 77 K.  The resulting coverage consisted of small 2D molecular clusters ranging from two to five molecules in size.  We therefore allowed the sample to warm to 200 K to encourage sufficient molecular diffusion to create large enough islands for study.  The resulting STM images and line profiles can be seen in Figure \ref{fig:Fig2}(a), where we clearly observe the C$_{60}$ fullerenes located at a single height above the Cu(111) surface with a measured height close to 670 $\pm$ 50 pm.  

The consistency in molecular height was further examined with sequentially acquired constant height STM and ncAFM images.  The switch from constant current STM mode to constant height ncAFM was carried out as follows.  After performing constant current STM, the feedback was paused and the probe withdrawn by approximately 30 nm, such that no force is felt by the AFM tip.  A frequency sweep was measured to establish free resonant frequency (f$_{0}$), after which the oscillation and PLL control was activated.  The probe was then slowly approached towards the surface in the constant height mode until contrast in the frequency shift channel was observed.  ncAFM scan heights, where relevant, are referenced against $\delta f(z)$ measurements. Constant height data, shown in Figure \ref{fig:Fig2}(b-d), were collected at 5 K, using C$_{60}$ terminated tips in order to precisely probe the intermolecular interaction \cite{Sweetman2016,Jarvis2015a}.  The resulting ncAFM frequency shift images show no detectable variation in molecular height, either in the attractive (c) or repulsive (c,d) imaging modes,  confirming the apparent absence of surface reconstruction in samples prepared at low temperature.  Moreover, despite a consistent height in ncAFM, the same molecular islands imaged with STM show that molecular rotation is still present.  

The observation of bright and dark hexagon-up C$_{60}$ fullerenes at room temperature, and the suppression of height variation at low temperatures clearly suggest that rotational effects alone are insufficient to explain the height variation observed in C$_{60}$ islands\cite{Tang2011,Shin2014,Gardener2009,Pabens2015,Colton1993}.  We therefore suggest that vacancy reconstruction must be present in this system, which we now quantify with ncAFM.


\begin{figure}[!htbp]
	\includegraphics[width=6cm]{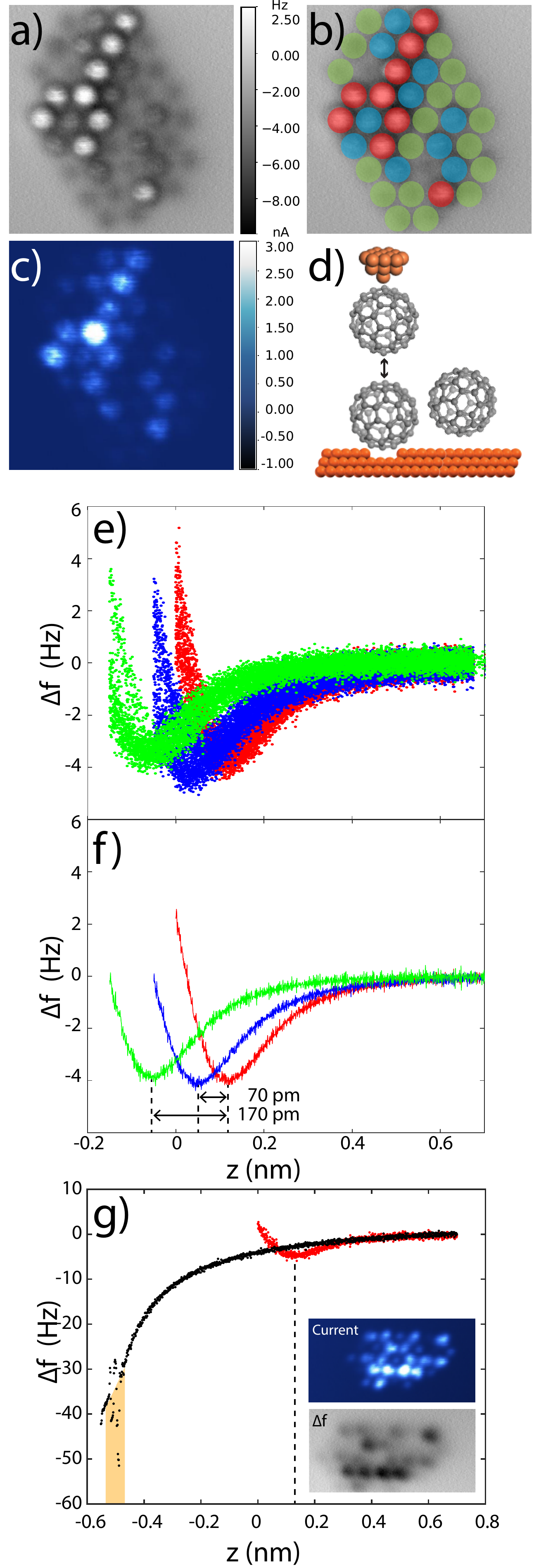}
	\caption{\label{fig:Fig3} \textbf{Relative height measurement with C$_{60}$ terminated probes. } \textbf{(a)} Constant height frequency shift image of a reconstructed C$_{60}$ island on Cu(111).  \textbf{(b)} Frequency shift image from (a) colour coded by molecular height. \textbf{(c)} Constant height current image obtained prior to (a).  \textbf{(d)} Schematic representation of the C$_{60}$-C$_{60}$ tip-sample $\Delta f(z)$ measurement. \textbf{(e)} Plots of the short-range chemical $\Delta f(z)$ for 24 of the molecules probed within the molecular island.  z=0 corresponds to the image height in (a). \textbf{(f)} Averaged plots of the short-range chemical $\Delta f(z)$ shown in (e). (g) Frequency shift plots showing short-range subtracted $\Delta f(z)$  collected over a `bright' C$_{60}$ molecule (red) and long range $\Delta f(z)$ collected above the Cu(111) surface (black).  The long-range black curve has been offset in y to aid comparison with the short range data.  The tip-sample contact point region is highlighted in orange.  Inset panels show constant height current and frequency shift images with the spectra locations marked as red and black stars. Parameters: (a,e,f) a$_0$=300 pm, V = -2.3 mV (c) a$_0$=300 pm, V = +1 V.  (g) Insets: a$_0$=300 pm, V=-1.3 mV.
	}
\end{figure}

\textbf{Direct height measurement with ncAFM}.  In order to quantify the relative heights of individual molecules within reconstructed C$_{60}$ islands we carried out site specific $\Delta f(z)$ spectroscopy measurements using C$_{60}$ terminated probes. Similar to force-mapping studies using CO terminated probes\cite{Schuler2013,Liu2016,Albrecht2016}, the C$_{60}$ termination provides a consistent tip-sample force interaction such that relative height measurements are precise to within tens of pm \cite{Chiutu2012,Sweetman2016}.  Tips were functionalised either serendipitously or by light contact with a molecule located at the edge of an island (see, for instance, the removal of a molecule in Figure \ref{fig:Fig2}(c)).  C$_{60}$ termination was confirmed by measuring the characteristic C$_{60}$ - C$_{60}$ interaction potential \cite{Chiutu2012}.  Short range frequency shift curves were extracted using the on-off island subtraction method\cite{Sweetman2014}, allowing quantitative height determination using the $\Delta f(z)$ turnaround point as a well defined reference.  Reconstructed fullerene islands were created by allowing the low temperature samples studied in Figure 2 to warm up to room temperature, before being cooled back down to 5 K for imaging.  

Constant height images of frequency shift and current are shown in Figure \ref{fig:Fig3} (a) and (c) respectively.  Both images show clear variation in brightness, confirming that molecular reorganisation has taken place.  The ncAFM image in particular shows significant variation, with some molecules appearing bright (i.e. positive $\Delta f(z)$ resulting from strong repulsion at close tip-sample separations), and other molecules appearing dark (negative $\Delta f(z)$ resulting from attractive tip-force interaction at larger tip-sample separations).  

$\Delta f(z)$ spectroscopy measurements were carried out for each molecule within the island using C$_{60}$ terminated probes (see cartoon in Figure \ref{fig:Fig3}(d)), resulting in the short-range $\Delta f(z)$ data shown in Figure \ref{fig:Fig3}(e).  Analysis of the $\Delta f(z)$ measurements reveal consistent groupings with turnarounds at three specific values of Z.  For clarity, the three groupings of $\Delta f(z)$ curves are marked in Figure \ref{fig:Fig3}(b) and have been plotted as red (8 curves), blue (9 curves), and green (7 curves) data points in Figure \ref{fig:Fig3}(e). Averaged curves are also shown in Figure \ref{fig:Fig3}(f). Using the $\Delta f(z)$ minima as a reference, the height difference between C$_{60}$ molecules is measured as 70 pm and 170 pm for blue and green $\Delta f(z)$ curves respectively, referenced to the position of the red $\Delta f(z)$ measurements.


An estimate for the absolute height of C$_{60}$ was obtained by comparing $\Delta f(z)$ spectra taken above the molecular island to $\Delta f(z)$ measured when contacting the Cu(111) surface.  The resulting spectra are shown in Figure \ref{fig:Fig3}(g), collected  above a `bright' (red) C$_{60}$ fullerene and off the island over clean Cu(111) (black).  Examination of the Cu(111) $\Delta f(z)$ reveals a clear cut-off point at which the tip-sample interaction becomes unstable, which we attribute to the onset of chemical binding between the C$_{60}$-terminated tip and the Cu(111) surface.  Although the onset of C$_{60}$-Cu(111) tip-surface chemical binding is clearly not as well-defined compared to the C$_{60}$ - C$_{60}$ tip-molecule potential, taking this as a best estimate of the Cu(111) surface height, we determine the C$_{60}$ height to be 650 $\pm $50 pm.  


The STM and ncAFM measurements provide clear evidence for substantial surface reconstruction for C$_{60}$ on Cu(111).  Both STM and ncAFM clearly identify three distinct molecular adsorption heights.  In the case of STM a fourth height is observed, however, as this small 20 pm variation is not observed with ncAFM, this is most likely attributed to molecular rotation. STM line profile and ncAFM $\Delta f(z)$ height measurements are summarised in Table 1. The relative differences observed between STM and ncAFM are attributed to rotational effects arising from enhanced tunnelling into the LUMO orbital in STM.   

\begin{table}[ht] 
\caption{Summary of C$_{60}$ molecular heights as measured by STM and ncAFM.  } 
\centering      
\begin{tabular}{c c c c}  
\hline\hline                        
 & Site 1 & Site 2 & Site 3 \\ [0.5ex] 
\hline                    
STM & 670 pm & 550 pm & 500 pm  \\    
ncAFM & 650 pm & 580 pm & 480 pm  \\  [1ex]       
\hline     
\end{tabular} 
\label{table:nonlin}  
\end{table}


Based on the ncAFM and STM results above, and comparisons to simulations as detailed below, we suggest the following assignments for the observed C$_{60}$ adsorption configurations.  The measured heights for the `bright' (red) C$_{60}$ fullerenes is consistent with the low temperature preparation, and so are assigned to C$_{60}$ adsorbed on unreconstructed regions of the surface, most likely adsorbed on hollow sites \cite{Larsson2008}.  `Middle' (blue) fullerenes have a height 70 pm below that of the bright molecules, suggesting the presence of a vacancy reconstruction.  Finally, the `dark' (green) fullerenes exhibit a measured height 170 pm below the bright molecules, suggesting the removal of an entire  layer of copper atoms, potentially indicative of the 7-atom nanopit model proposed by LEED I(V) analysis \cite{Pai2010,Xu2012}.  These assignments are consistent with simulations of the nanopit and vacancy structures.  The nanopit structure was investigated in detail \cite{Pai2010,Xu2012} using density functional theory simulations which suggested that adsorption required removal of the top Cu layer.  We note that a single Cu layer is 209 pm \cite{Straumanis1969} in height, which agrees well with our measured height difference between the `bright' and `dark' C$_{60}$ adsorption sites.   In contrast, the vacancy structure results in much smaller height differences, making it clearly distinguishable from the nanopit structure.  Calculations using DFT approaches typically results in a height reduction up to 50 pm \cite{Li2009}, relative to top-layer surface adsorption, similar to our measured height difference between the `bright' and `middle' C$_{60}$ adsorption sites.  Indeed, a study on Pt(111)\cite{Pinardi2014} suggests that vacancy site reconstruction is complicated, with the location of the ejected vacancy atom apparently playing an important role in the exact adsorption height of the C$_{60}$.

In summary, combined STM and ncAFM with single molecule $\delta f(z)$ spectroscopy provides direct evidence for complex height variation in C$_{60}$:Cu(111) prepared at room temperature, suggesting a mixture of adsorption sites consistent with models of non-reconstructed, vacancy, and nanopit C$_{60}$ adsorption configurations.   Furthermore, we show that this reconstruction can be suppressed by low temperature preparation.  We suggest that rather than a single preferred adsorption site, a mixture of adsorption configurations may be present, and indeed prevalent, not only on Cu(111) but also other C$_{60}$ metal (111) systems with similar lattice constant, providing an explanation for the observed bright and dark features in STM.  The complexity of this surface reconstruction is essential to understand not only to the surface geometry, but also electronic and chemical properties important for applications of fullerene surface chemistry.


\bibliography{C60Cu111Vac_1}

\end{document}